\documentclass[a4paper,11pt]{article}
\usepackage{pos}

\newcommand{\als}{\alpha_{\rm s}}
\def\nf{{n^{}_{\! f}}}
\def\floo{fl_{11}}

\title{
\vspace*{-2.3cm}
\begin{minipage}{\textwidth}
{\normalfont\small DESY 22-139, Nikhef 2022-011, LTH 1311
\hspace{\fill} August 2022}\\
\end{minipage}\\[42pt]
DIS coefficient functions at four loops in QCD and beyond
}
\ShortTitle{DIS coefficient functions at four loops in QCD and beyond}

\author*[a]{S. Moch}
\author[b]{B. Ruijl}
\author[c]{T. Ueda}
\author[d]{J.A.M. Vermaseren}
\author[e]{A. Vogt}

\affiliation[a]{
  II. Institut f\"ur Theoretische Physik, 
  Universit\"at Hamburg, \\
  Luruper Chaussee 149, 
  22761 Hamburg, Germany}

\affiliation[b]{
  ETH Z\"urich\\
  R\"amistrasse 101, 8092 Z\"urich, Switzerland}

\affiliation[c]{
  Faculty of Science and Technology, Seikei University\\ 
  Musashino, Tokyo 180-8633, Japan}

\affiliation[d]{
  NIKHEF Theory Group, \\
  Science Park 105, 1098 XG Amsterdam,
  The Netherlands}

\affiliation[e]{
  Department of Mathematical Sciences, University of Liverpool,\\
  Liverpool L69 3BX, UK}

\emailAdd{sven-olaf.moch@desy.de}
\emailAdd{benruyl@gmail.com}
\emailAdd{tueda@st.seikei.ac.jp}
\emailAdd{t68@nikhef.nl}
\emailAdd{Andreas.Vogt@liv.ac.uk}

\abstract{
We report results for the lowest even-$N$ moments of the 
flavor-nonsinglet structure functions $F_2$ and $F_L$ in QCD at the fourth order 
in the perturbative expansion in the strong coupling constant $\alpha_s$.
Our results are presented in numerical form and we compare them with the
leading and subleading terms of the threshold expansion for large values of $N$, 
which corresponds to the limit $x \to 1$.
}

\FullConference{%
  Loops and Legs in Quantum Field Theory - LL2022,\\
  25-30 April, 2022\\
  Ettal, Germany
}

\begin{document}
\maketitle

\section{Introduction}

The deep-inelastic scattering (DIS) of leptons off a nucleon target 
is one of the basic scattering reactions in Quantum Chromodynamics (QCD), 
which has been studied in many collider experiments during the past 50 years.
In recent times high precision inclusive DIS data has been measured at HERA~\cite{Abramowicz:2015mha} and 
in the future the forthcoming Electron-Ion-Collider (EIC)~\cite{Boer:2011fh,AbdulKhalek:2021gbh} 
is expected to contribute as well.
Unpolarized inclusive lepton-nucleon DIS proceeds through the reaction
\begin{equation}
\label{eq:dis}
  l(k) \:+\: {\rm nucl}(p) \:\:\rightarrow\:\: l^{\,\prime}(k^{\,\prime}) \:+\:  X
  \, ,
\end{equation}
where the scattered leptons (or neutrinos) are denoted by $l$ and $l^{\,\prime}$ 
and the nucleon state by `nucl', with the respective momenta $k$,
$k^{\,\prime}$ and $p$, and $X$ is the inclusive hadronic final state.  
The charge of the exchanged gauge boson $V(q)$ with momentum $q = k-k^{\,\prime}$, 
is used to classify neutral- ($V = \gamma^*, Z$) or charged-current ($V = W^\pm$) DIS.
The theoretical predictions for the reaction~(\ref{eq:dis}) are based on the well known structure
functions $F_i$, with $i=1,2,3$ and $F_L=F_2-2xF_1$, which are functions of
the Bjorken $x$ variable $x = Q^2 / (2p \cdot q)$, where $0 < x \leq 1$, 
and the scale $Q^2=-q^2>0$ of the exchanged virtual boson.

Perturbative QCD allows for the computation of the scale dependence of the 
structure functions as a series in the strong coupling $\alpha_s$, 
together with the coefficient functions of the hard scattering process.
The current state-of-the art for massless perturbative QCD
is the next-to-next-to-next-to-leading order approximation (abbreviated as (next-to)$^3$-leading order or N$^3$LO) 
for the structure functions $F_i$, with $i=1,2,3$.
This encompasses the complete three-loop coefficient functions for neutral-~\cite{Vermaseren:2005qc} 
and charged-current~\cite{Moch:2007gx,Moch:2007rq,Moch:2008fj} DIS 
and partial results for the four-loop splitting functions 
(Mellin moments~\cite{Velizhanin:2011es,Velizhanin:2014fua,Ruijl:2016pkm,Moch:2017uml,Moch:2018wjh,Moch:2021qrk}, 
the large-$\nf$ contributions~\cite{Gracey:1994nn,Davies:2016jie}, 
and the planar limit of the nonsinglet case~\cite{Moch:2017uml}) governing the $Q^2$ dependence, 
so that the N$^3$LO predictions are robust within the kinematic range of past
(HERA) and future (EIC) DIS experiments.

For theory predictions beyond this order, i.e., at the (next-to)$^4$-leading order (N$^4$LO), 
it is necessary to consider the coefficient functions at four loops in perturbative QCD, which is ongoing
work and will be reported in these proceedings.

\section{Computation}

The focus in these proceedings is on the structure functions $F_2^{\gamma^*}$ and $F_L^{\gamma^*}$, 
which describe the one-photon exchange in neutral-current DIS.
The standard QCD factorization in leading twist approximation, i.e., disregarding terms suppressed by powers of $1/Q^2$,
allows to express them as convolutions of the coefficient functions with the parton distributions (PDFs), 
\begin{equation}
\label{eq:Fns-cq}
  F^{\gamma^*}_a(x,Q^2)
  \, = \, 
  \left[ C_a(Q^2) \otimes q_{+,\rm ns}^{\,}(Q^2) \right] \! (x)
  \, ,
  \qquad
  a = 2,\,  L
  \, ,
\end{equation}
where $q_{+,\rm ns}$ is the nonsinglet quark PDF in the nucleon, 
the dependence on the factorization scale $\mu^2$ is suppressed, and
$\otimes$ denotes the standard convolution. 
The coefficient functions in Eq.~(\ref{eq:Fns-cq}) 
have an expansion in powers of the strong coupling $a_{\rm s} = \alpha_{\rm s}/(4\pi)$, 
\begin{equation}
\label{eq:cf-exp}
  C_{a,{\rm ns}} \, = \, \delta_{a2}\, + \sum_{l=1}^{\infty} \, a_{\rm s}^{\, l} c_{a,{\rm ns}}^{\,(l)} 
  \, ,
  \qquad
  a = 2,\,  L
  \, ,
\end{equation}
where we have suppressed the dependence on $x$, 
on the number of effectively massless flavors $\nf$ and on the ratio of scales $Q^2/\mu^2$.
Their Mellin moments are defined as 
\begin{equation}
\label{eq:Mellin-def}
  C_{a,{\rm ns}}(N) \,=\, \int_0^1 \! dx\; x^{\,N-1}\, C_{a,{\rm ns}}(x)
  \, ,
  \qquad
  a = 2,\,  L
  \, ,
\end{equation}
and the even ones are accessible within the framework of the operator product expansion (OPE)  
by computing suitable projections of the imaginary part of the forward Compton amplitude 
for the scattering process of a virtual photon off a quark. 
The computational set-up, building on the OPE, is well established~\cite{Buras:1979yt}, 
and has been used and described in detail in previous publications, 
see e.g., Refs.~\cite{Moch:2004pa,Vogt:2004mw,Vermaseren:2005qc,Moch:2007gx,Moch:2007rq,Moch:2008fj,Ruijl:2016pkm}.

\begin{figure}[tb]
  \begin{center}    
    \includegraphics[width=0.35\textwidth, angle=0]{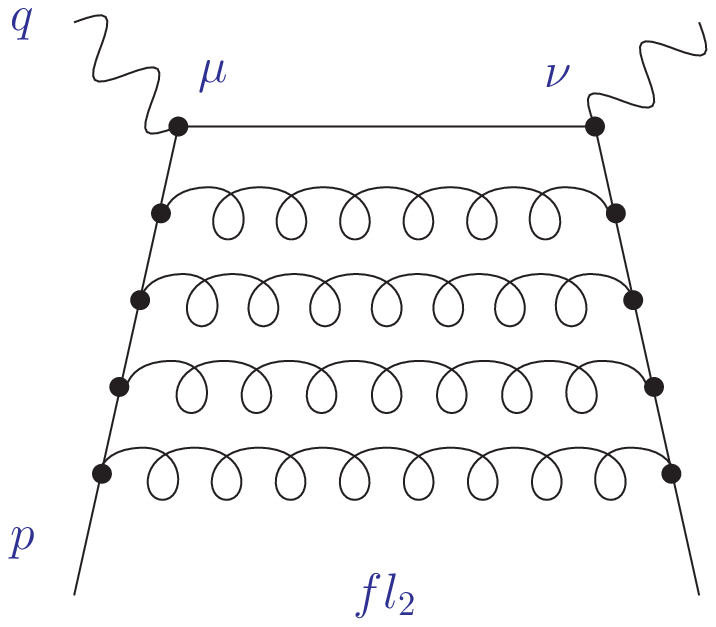}
    \hspace*{15mm}
    \includegraphics[width=0.35\textwidth, angle=0]{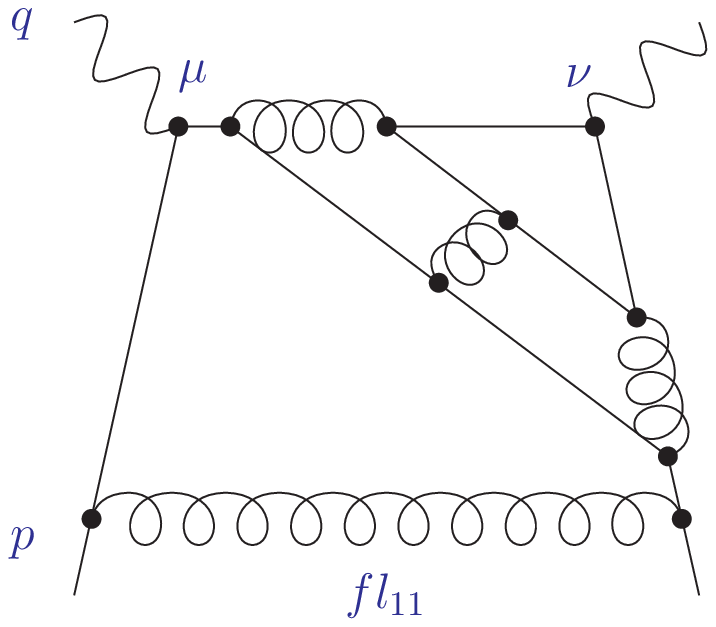}
  \end{center}
\caption{\small 
\label{fig:flavor-factors}
Typical Feynman diagrams for the different nonsinglet flavor topologies, 
$fl_{2}$ (left) and $fl_{11}$ (right).
The latter admits a $d_F^{abc}d_F^{abc}$ color factor starting from three loop order through the coupling of three gluons.
}
\end{figure}

All contributing Feynman diagrams for the process 
\begin{equation}
\label{eq:partonic-dis}
  \gamma^*(q) \:+\: q(p) \:\:\rightarrow\:\: X
  \, ,
\end{equation}
up to four loops are generated with {\sc Qgraf}~\cite{Nogueira:1991ex} 
and their color coefficients are obtained for a general $SU(n_c)$ gauge group 
using the algorithms of~\cite{vanRitbergen:1998pn}, which determine the group invariants for (semi-)simple Lie groups.
At four loops, the color factors are given by powers of the quadratic Casimirs 
$C_F^{4-k}C_A^k$ for $k=0,\dots, 3$ as well as by 
combinations of the cubic ones $C_F d_{F}^{\,abc}d_{F}^{\,abc}$ and $C_A d_{F}^{\,abc}d_{F}^{\,abc}$ 
and the quartic ones $d_{F}^{\,abcd}d_{A}^{\,abcd}$ and $d_{F}^{\,abcd}d_{F}^{\,abcd}$, 
the latter being obtained from the symmetrized trace $d_{r}^{\,abc}$ ($d_{r}^{\,abcd}$) of three (four) $SU(n_c)$
generators in representation $r$ of the $SU(n_c)$. 
In addition, $\nf$-dependent terms arise by replacing $C_F \to \nf$ for up to three powers of $C_F$. 
Besides their color factors diagrams are grouped according to their nonsinglet flavor topologies. 
For neutral-current diagrams up to four loops the corresponding charge factors
read~\cite{Larin:1996wd,Vermaseren:2005qc}
\begin{equation}
  \label{eq:flavor-nc}
  fl_{2} \,=\, 1\, ,
  \qquad\qquad\qquad
  fl_{11} \,=\, 3 \langle e \rangle\, 
  \,=\, \frac{3}{\nf}\:\sum_{i=1}^{\nf}\, e_{q^{}_i}\, ,
\end{equation}
where $\langle e \rangle$ represents the average of the quark charges $e_{q^{}_i}$.
Thus, for $\nf = \{1, \dots, 6\}$ the coefficient 
$fl_{11}$ takes the numerical values $fl_{11}=\{-1, 1/2, 0, 1/2, 1/5, 1/2\}$ 
and diagrams of the $fl_{11}$ flavor topology arise only with cubic color factors $d_{F}^{\,abc}d_{F}^{\,abc}$.
An illustration of the nonsinglet flavor topologies at four loops is given in Fig.~\ref{fig:flavor-factors}.

The matrix elements for the process~(\ref{eq:partonic-dis}) contain divergencies 
at higher orders, which are regularized in $D=4-2\epsilon$ dimensions.
Choosing a fixed Mellin moment $N$ as defined in Eq.~(\ref{eq:Mellin-def}) 
then reduces the expressions for the Feynman diagrams to two-point functions 
up to four-loop order, i.e., massless propagator-type diagrams, 
which can be computed with the help of 
standard integration-by-parts algorithms~\cite{Tkachov:1981wb,Chetyrkin:1981qh}, 
whose solutions are encoded in the program {\sc Forcer}~\cite{Ruijl:2017cxj} 
for the computer algebra system {\sc Form}~\cite{Vermaseren:2000nd,Kuipers:2012rf,Ruijl:2017dtg}
and its multi-threaded version {\sc TForm}~\cite{Tentyukov:2007mu}, 
which is used for all symbolic manipulations.

This computation provides an analytic result for the bare forward Compton
amplitude at a chosen fixed value of $N$ which is then subject to the standard ultraviolet renormalization of the 
strong coupling $\alpha_{\rm s}$ and the subsequent removal 
of the remaining collinear singularities in powers of $1/\epsilon$ 
in the minimal subtraction scheme \cite{tHooft:1973mfk}. 
The single poles in $1/\epsilon$ provide the anomalous dimensions of the 
quark nonsinglet operator matrix elements~\cite{Moch:2017uml}.

High values of $N$, however, give rise to high powers 
of both, propagators in the denominator and scalar products in the numerator 
in the propagator-type loop diagrams, to be computed with {\tt Forcer}.
This leads to large-size intermediate expressions, sometimes of the order of several TByte and to long run times.
For example, the four-loop result for the Mellin moment $N=10$ of the coefficient function $C_{L,{\rm ns}}$ 
in Eq.~(\ref{eq:cf-exp}) required the evaluation ${\cal O}(3200)$ Feynman diagrams 
with intermediate expressions of sizes up to  ${\cal O}(20)$ TByte 
and a total of ${\cal O}(800000)$ hrs CPU time.
The multi-threaded version {\sc TForm} delivers an average speed-up factor of ${\cal O}(10)$.

\section{Coefficient functions at four loops}

We can now provide the exact even Mellin moments for $N=2,\, 4,\, 6,\, 8,\, 10$ 
of the coefficient functions $C_{2,{\rm ns}}$ and $C_{L,{\rm ns}}$ 
up to fourth order in $\als$ for QCD, 
i.e., taking the numerical values of the $SU(n_c)$ color coefficients for $n_c=3$.

The complete perturbative expansion for $C_{2,\rm ns}(N)$ for $\nf=4$ reads 
\begin{eqnarray}
\label{eq:c2numbers}
  C_{2,\rm ns}(2) &=&
  1 +0.0354\,\,\als-0.0231\,\,\als^2-0.0613\,\,\als^3-0.4746\,\,\als^4
\nonumber\\ && \qquad
  +\floo\*\left( -0.0486\,\,\als^3-0.1424\,\,\als^4 \right)
\, ,
\nonumber\\[0.2mm]
  C_{2,\rm ns}(4) &=&
  1 +0.4828\,\,\als+0.4711\,\,\als^2+0.4727\,\,\als^3-0.2458\,\,\als^4
\nonumber\\ && \qquad
  +\floo\*\left( -0.0367\,\,\als^3-0.0893\,\,\als^4 \right)
\, ,
\nonumber\\[0.2mm]
  C_{2,\rm ns}(6) &=&
  1 +0.8894\,\,\als+1.2054\,\,\als^2+1.7572\,\,\als^3+1.7748\,\,\als^4
\nonumber\\ && \qquad
  +\floo\*\left( -0.0325\,\,\als^3-0.0684\,\,\als^4 \right)
\, ,
\nonumber\\[0.2mm]
  C_{2,\rm ns}(8) &=&
  1 +1.2358\,\,\als+2.0208\,\,\als^2+3.5294\,\,\als^3+5.3921\,\,\als^4
\nonumber\\ && \qquad
  +\floo\*\left( -0.0304\,\,\als^3-0.0591\,\,\als^4 \right)
\, ,
\nonumber\\[0.2mm]
  C_{2,\rm ns}(10) &=&
  1 +1.5359\,\,\als+2.8608\,\,\als^2+5.6244\,\,\als^3+10.324\,\,\als^4
\nonumber\\ && \qquad
  +\floo\*\left( -0.0291\,\,\als^3-0.0547\,\,\als^4 \right)
\, ,
\end{eqnarray}
and for $C_{L,\rm ns}(N)$ for $\nf=4$
\begin{eqnarray}
\label{eq:cLnumbers}
  C_{L,\rm ns}(2) &=&
  0.14147\,\,\als \*\left(1 +1.7270\,\,\als+3.7336\,\,\als^2+9.5619\,\,\als^3
\right. \nonumber\\ && \qquad \left.
  +\floo\*\left(
    -0.1102\,\,\als^2-0.7865\,\,\als^3\right)
\right)
\, ,
\nonumber\\[0.2mm]
  C_{L,\rm ns}(4) &=&
0.08488\,\,\als \*\left(1 +2.5619\,\,\als+6.9208\,\,\als^2+20.251\,\,\als^3
\right. \nonumber\\ && \qquad \left.
  +\floo\*\left(
-0.1201\,\,\als^2-0.9983\,\,\als^3\right)
\right)
\, ,
\nonumber\\[0.2mm]
  C_{L,\rm ns}(6) &=&
0.06063\,\,\als \*\left(1 +3.1557\,\,\als+9.6370\,\,\als^2+30.573\,\,\als^3
\right. \nonumber\\ && \qquad \left.
  +\floo\*\left(
-0.1232\,\,\als^2-1.1174\,\,\als^3\right)
\right)
\, ,
\nonumber\\[0.2mm]
  C_{L,\rm ns}(8) &=&
0.04716\,\,\als \*\left(1 +3.6191\,\,\als+12.040\,\,\als^2+40.535\,\,\als^3
\right. \nonumber\\ && \qquad \left.
  +\floo\*\left(
-0.1245\,\,\als^2-1.1997\,\,\als^3\right)
\right)
\, ,
\nonumber\\[0.2mm]
  C_{L,\rm ns}(10) &=&
0.03858\,\,\als \*\left(1 +4.0020\,\,\als+14.215\,\,\als^2+50.164\,\,\als^3
\right. \nonumber\\ && \qquad \left.
  +\floo\*\left(
-0.1253\,\,\als^2-1.2629\,\,\als^3\right)
\right)
\, ,
\end{eqnarray}
where the numerical values for $C_{2,{\rm ns}}$ and $C_{L,{\rm  ns}}$ (flavor class $fl_{2}$ only) at $N=2,\, 4,\, 6$ 
have already been quoted in~\cite{Ruijl:2016pkm}.
Further results for $C_{2,{\rm ns}}$ and $C_{L,{\rm ns}}$ at $N=12$ and $N=14$
have also been obtained, but are limited to the large-$n_c$ approximation.
All Mellin moments agree with the published all-order large-$\nf$ limits of the nonsinglet structure functions 
$F_2$~\cite{Mankiewicz:1997gz} and $F_L$~\cite{Gracey:1995aj}.

In addition to those moments more information about the four-loop coefficient functions 
is available in the kinematic limits of large $N$ ($x \to 1$) and small $N \simeq 0$ ($x \to 0$). 
Near threshold the coefficient function $c_{2,{\rm ns}}^{\,(4)}$ is dominated 
by singular plus-distributions $[\ln^k(1-x)/(1-x)]_+$ with $7 \geq k \geq 0$ for $x \to 1$ 
($\als^n \ln^{k+1}(N)$ in Mellin-$N$ space)
and the coefficients of all terms with $k \geq 1$ are known, see, e.g.~\cite{Das:2019btv}. 
For $n_c=3$ in QCD $c_{2,{\rm ns}}^{\,(4)}$ reads 
\begin{eqnarray}
\label{eq:c2dis4-D0}
{\lefteqn{
  c^{(4)}_{2, {\rm ns}}(x) \,=\, 
  16.85596708\*\left[ \frac{\ln^{\,7} (1-x)}{1-x} \right]_+ 
  + \ldots }}
\nonumber \\
&& 
  + \bigl\{
  (3.88405 \pm 0.1)\cdot 10^4
  + (-3.49648951 \pm 0.00000004)\cdot 10^4\, \*\nf 
\nonumber \\
&& 
  + 2062.715183\, \*\nf^{\!\! 2}
  - 12.08488248\, \*\nf^{\!\! 3}
  + 47.55183089\, \*\nf \* \floo 
    \bigr\}\*\left[ \frac{1}{1-x} \right]_+
\nonumber \\
&& 
    +\delta(1-x)\*c_{4, \delta, {\rm DIS}}
\nonumber \\
&& 
    - 16.85596708\*\ln^7(1-x) 
    + \bigl\{
      504.6255144
    - 14.74897119\, \*\nf
      \bigr\}\*\ln^6(1-x) 
\nonumber \\
&& 
    + \bigl\{
    - 5135.705824
    + 416.4828532\, \*\nf
    - 4.213991769\, \*\nf^{\!\! 2}
      \bigr\}\*\ln^5(1-x) 
\nonumber \\
&& 
    + \bigl\{
      20935.61036
    - 4034.293546\, \*\nf
    + 108.5761316\, \*\nf^{\!\! 2}
    - 0.3950617283\, \*\nf^{\!\! 3}
      \bigr\}\*\ln^4(1-x) 
\nonumber \\
&& 
    +{\cal O}(\ln^3(1-x))
\, ,
\end{eqnarray}
where all exact values have been rounded to ten digits and 
the flavor factor $fl_{11}$ is given in Eq.~(\ref{eq:flavor-nc}). 
The new result for the term proportional to $[1/(1-x)]_+$ 
is based on the Mellin moments up to $N \leq 15$ 
for $C_{2,{\rm ns}}$ and $C_{3,{\rm ns}}$. 
The latter denotes the coefficient function of the 
charged-current structure function $F_3^{W^+ + W^-}$, 
with its odd Mellin moments being accessible in the OPE.
The limits $x \to 1$ for $C_{2,{\rm ns}}$ and $C_{3,{\rm ns}}$ coincide, 
thus increasing the available information for the estimates in Eq.~(\ref{eq:flavor-nc}). 
Also, unpublished results for much higher Mellin moments (up to 
$N \simeq 40$) for the $\nf$-dependent contributions have been used.
The coefficient of $[1/(1-x)]_+$ in Eq.~(\ref{eq:c2dis4-D0}) 
thus supersedes the previous estimate~\cite{Das:2019btv}, 
where also numerical estimates for the term $c_{4, \delta, {\rm DIS}}$
proportional to $\delta(1-x)$ have been presented.
The logarithmically enhanced powers $\ln^k(1-x)$ for $7\geq k \geq 4$ 
of the first subleading corrections at power $(1-x)^0$ 
($\als^n \ln^{k}(N)/N$ in Mellin-$N$ space) for $c_{2,{\rm ns}}^{\,(4)}$ 
in Eq.~(\ref{eq:c2dis4-D0}) have been derived in~\cite{Moch:2009hr}, 
with unknown terms starting at ${\cal O}(\ln^3(1-x))$, as indicated 
(see also \cite{Ajjath:2020sjk} for recent work).
For $c_{L,{\rm ns}}^{\,(4)}$ they are actually leading and the 
expansion around the limit $x \to 1$ for $n_c=3$ in QCD reads
\begin{eqnarray}
\label{eq:cldis4-D0}
  c^{(4)}_{L, {\rm ns}}(x) &=& 
      16.85596708\*\ln^6(1-x) 
    + \bigl\{
          - 306.0638892
          + 21.06995884\, \*\nf
      \bigr\}\*\ln^5(1-x) 
\nonumber \\
&& 
    + \bigl\{
            2421.032535
          - 356.9371659\, \*\nf
          + 9.481481481\, \*\nf^{\!\! 2}
      \bigr\}\*\ln^4(1-x) 
\nonumber \\
&& 
    + {\cal O}(\ln^3(1-x))
\, .
\end{eqnarray}
The small-$x$ behavior of the nonsinglet coefficient functions features 
double logarithms in the limit $x \to 0$, which appear as 
$\als^n \ln^{k}(x)$, where $k=2n-1, \dots, 1$, at all orders 
($\als^n/N^{k+1}$ in Mellin-$N$ space for $N \to 0$), 
which have been resummed to the third logarithmic (N$^2$LL) order~\cite{Vogt:2012gb,Davies:2022ofz}.
The fixed-order expansions of the latter resummations gives 
for $c^{(4)}_{2, {\rm ns}}$ and $c^{(4)}_{L, {\rm ns}}$ the following results~\cite{Davies:2022ofz},
\begin{eqnarray}
\label{eq:c2smallx}
  c^{(4)}_{2, {\rm ns}}(x) &=&
  - \frac{13}{168}\* C_F^4\* \ln^{7}(x)
  + \Bigg\{
    \frac{263}{180} \*C_F^4 - \frac{911}{1080} \*C_F^3\*\beta_0 
  \Bigg\} \* \ln^{6}(x)
  - \Bigg\{ 
   \left(\frac{14}{5} - \frac{734}{15} \* \zeta_2 \right) \*C_F^4
  - \frac{56}{15}\* C_F^3\*\beta_0
\nonumber\\
&& 
  + \left(\frac{109}{18} + \frac{208}{5} \* \zeta_2 \right) \*C_F^3 \* C_A
  - 13 \* \zeta_2 \*C_F^2 \* C_A^2
  + \frac{1951}{720}\* C_F^2\*\beta_0^{2}  
  \Bigg\} \* \ln^{5}(x)
+ {\cal O}\!\left(\ln^{4}(x)\right)
\, ,
\nonumber\\
  c^{(4)}_{L, {\rm ns}}(x) &=&
  - 2\* C_F^4\* \ln^{5}(x)
  + \Bigg\{
    \frac{59}{3} \*C_F^4 - \frac{124}{9} \*C_F^3\*\beta_0 
  \Bigg\} \* \ln^{4}(x)
    + \bigg\{
   \left(\frac{322}{3} + \frac{2008}{3} \* \zeta_2 \right) \*C_F^4
   - \frac{28}{3}\* C_F^3\*\beta_0
\nonumber\\
&& 
   - ( 80 + 640\*\zeta_2 ) \* C_F^3\*C_A
   + 200 \* \zeta_2 \*C_F^2 \* C_A^2
   - \frac{230}{9} \*C_F^2\*\beta_0^{2}
  \bigg\} \* \ln^{3}(x) 
+ {\cal O}\!\left(\ln^{2}(x)\right)
\, ,
\end{eqnarray}
which have been quoted for a general $SU(n_c)$ gauge theory.
While the partial results are of limited direct use for phenomenology,
i.e., still lacking knowledge on the logarithms 
$\ln^{k}(x)$ with $4 \geq k \geq 1$ for $c_{2,{\rm ns}}^{\,(4)}$ 
and $k = 1,2$ for $c_{L,{\rm ns}}^{\,(4)}$,
these expressions provide valuable information for checks of analytic computations 
or for reconstructions of the exact result from a number of fixed Mellin moments. 

\begin{figure}[t!]
\begin{center}
\includegraphics[width=0.97\textwidth, angle=0]{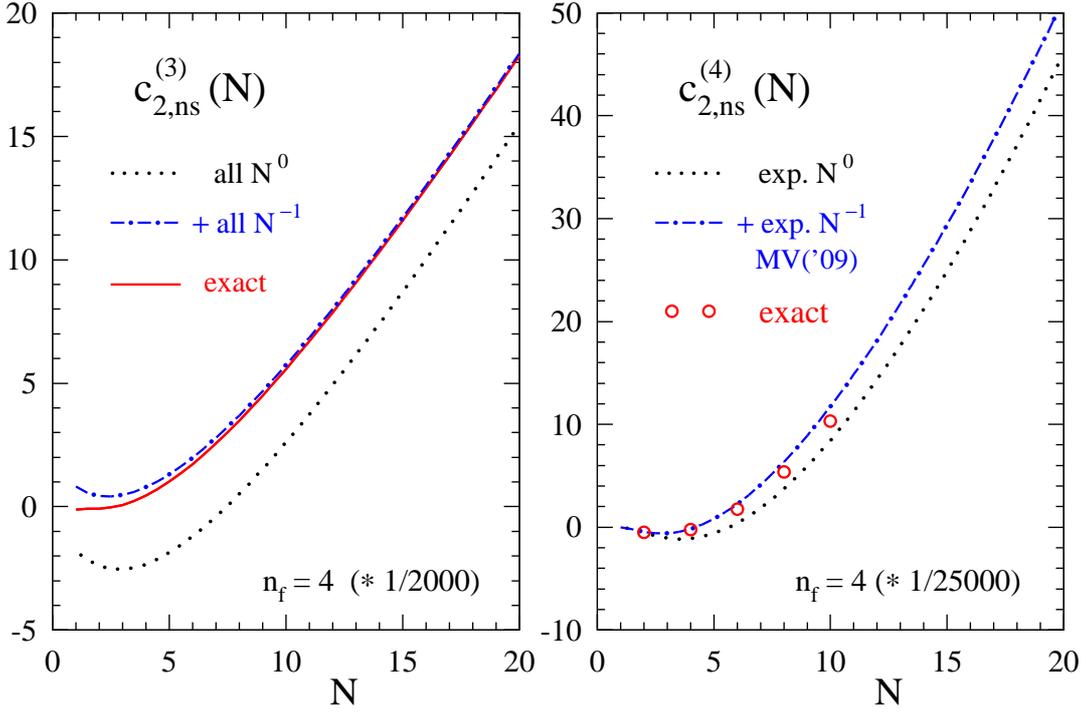}
\caption{
\label{fig:c2ns34nf4} 
The exact $N$-space results for the third-order coefficient function 
$c_{2,\rm ns}^{\,(3)}$ at $\nf=4$ for neutral-current (one-photon exchange)
DIS from~\cite{Vermaseren:2005qc} (left)  
and the moments in Eq.~(\ref{eq:c2numbers}) of the fourth-order term $c_{2,\rm ns}^{\,(4)}$ 
calculated so far (right).
In both cases also the contributions provided by large-$N$ resummations are shown.
}
\end{center}
\end{figure}
In Fig.~\ref{fig:c2ns34nf4} we plot the results for 
the coefficient function $C_{2,{\rm ns}}$ for $\nf=4$ at three and four loops 
as a function of the Mellin moment $N$.
The exact result for $c_{2,{\rm ns}}^{\,(3)}$ in Fig.~\ref{fig:c2ns34nf4} (left) 
is known from~\cite{Vermaseren:2005qc} and shown to lie in between the curves 
determined by the threshold logarithms $\als^3 \ln^{k}(N)$ 
and the first power suppressed terms $\als^3 \ln^{k}(N)/N$, 
with $6 \geq k \geq 1$ in both cases.
The new four-loop moments for $c_{2,{\rm ns}}^{\,(4)}$ in Fig.~\ref{fig:c2ns34nf4} (right) 
are also shown in comparison to the threshold logarithms, i.e., 
the $N$-space equivalent of Eq.~(\ref{eq:c2dis4-D0}) and including the additional leading $1/N$ enhanced terms.
Again, the two approximations are expected to bracket the, yet unknown, full result.

\begin{figure}[t]
\begin{center}
\includegraphics[width=0.97\textwidth, angle=0]{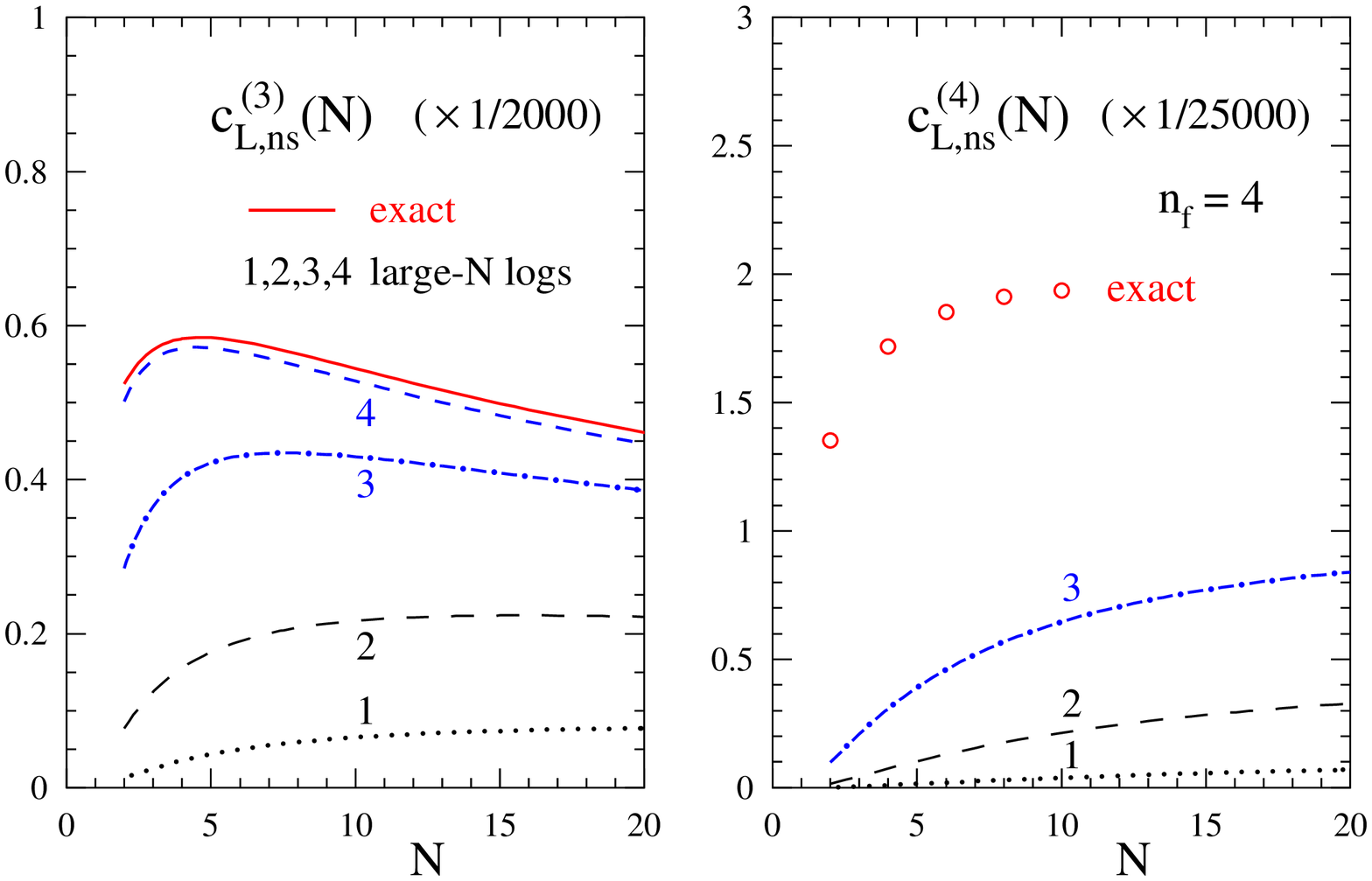}
\caption{
\label{fig:clns34nf4} 
Same as Fig.~\ref{fig:c2ns34nf4} for the exact results for $c_{L,\rm ns}^{\,(3)}$ from~\cite{Vermaseren:2005qc} (left)  
and the moments in Eq.~(\ref{eq:cLnumbers}) of $c_{L,\rm ns}^{\,(4)}$ (right).
}
\end{center}
\end{figure}
In Fig.~\ref{fig:clns34nf4} the same information at three and four loops is shown for 
the coefficient function $C_{L,{\rm ns}}$ for $\nf=4$. 
Here, the power suppressed terms proportional to $1/N$ are leading as $N \to \infty$.
From the three-loop result in Fig.~\ref{fig:clns34nf4} (left) it is obvious,
that the complete tower of logarithms $\als^3 \ln^{k}(N)/N$ with $4\geq k \geq
1$ is required approximate the exact result.
Therefore, it is evident, that at four loops the known first three logarithms 
$\als^4 \ln^{k}(N)/N$ with $6\geq k \geq 4$ for $c_{L,{\rm ns}}^{\,(4)}$
(given in Eq.~(\ref{eq:cldis4-D0}) in $x$-space) 
do not approximate the exact Mellin moments well and the further subleading
logarithms are needed as well.

\section{Summary and Outlook}

We have discussed the computation of the even Mellin moments $N=2, \dots, 14$ 
of the four-loop coefficient functions $C_{2,{\rm ns}}$ and $C_{L,{\rm ns}}$ 
in neutral-current (photon-exchange) DIS together with known results on the
endpoint behavior for large and small $x$ from all-order
considerations of soft and collinear dynamics or the high energy limit.
The results are available for a general $SU(n_c)$ gauge group and numerical
values for QCD ($n_c=3$) have been presented above.
Part of the ongoing studies are also the computation the 
odd moments $N=1, \dots, 15$ for those of the structure function $F_3$ in charged-current ($W^++W^-$-boson exchange) DIS, 
with the results for the highest moments $N=11$ to $15$ being restricted to the large-$n_c$ limit.
This agrees with and extends previous results on the first Mellin moment of $F_3$, 
i.e., the $\alpha_s^4$ contribution to the Gross-Llewellyn-Smith (GLS) sum rule~\cite{Baikov:2010je,Baikov:2012zn}.

The new results for the four-loop coefficient functions can be combined with
the low-$N$ moments of the five-loop nonsinglet anomalous
dimensions~\cite{Herzog:2018kwj} to allow for inclusive DIS predictions at N$^4$LO in massless QCD.
This situation is somewhat reminiscent to one more than 20 years ago, when the
then available three-loop moments of the DIS structure functions~\cite{Larin:1996wd,Retey:2000nq} 
were used to derive reliable approximations~\cite{vanNeerven:2000wp,vanNeerven:2001pe} 
in the kinematic range relevant for DIS experiments, 
before the computation of the exact three-loop splitting and coefficient
functions~\cite{Moch:2004pa,Vogt:2004mw,Vermaseren:2005qc}.

The combination of the new four-loop moments for $F_2$, $F_L$ and $F_3$ together
with information on the limits $x \to 0$ and $x \to 1$ 
will again provide robust approximations for very precise QCD predictions. 
They will be presented in a future publication~\cite{MRUVV-tba}.


\begin{thebibliography}{10}

\bibitem{Abramowicz:2015mha}
{\scshape H1, ZEUS} collaboration, H.~Abramowicz et~al., \emph{{Combination of
  measurements of inclusive deep inelastic ${e^{\pm }p}$ scattering cross
  sections and QCD analysis of HERA data}},
  \href{https://dx.doi.org/10.1140/epjc/s10052-015-3710-4}{\emph{Eur. Phys. J.
  C} {\bf 75} (2015) 580} [\href{https://arxiv.org/abs/1506.06042}{{\tt
  arXiv:1506.06042}}].

\bibitem{Boer:2011fh}
D.~Boer et~al., \emph{{Gluons and the quark sea at high energies:
  Distributions, polarization, tomography}},
  \href{https://arxiv.org/abs/1108.1713}{{\tt arXiv:1108.1713}}.

\bibitem{AbdulKhalek:2021gbh}
R.~Abdul~Khalek et~al., \emph{{Science Requirements and Detector Concepts for
  the Electron-Ion Collider: EIC Yellow Report}},
  \href{https://arxiv.org/abs/2103.05419}{{\tt arXiv:2103.05419}}.

\bibitem{Vermaseren:2005qc}
J.A.M.~Vermaseren, A.~Vogt and S.~Moch, \emph{{The Third-order QCD corrections
  to deep-inelastic scattering by photon exchange}},
  \href{https://dx.doi.org/10.1016/j.nuclphysb.2005.06.020}{\emph{Nucl. Phys.
  B} {\bf 724} (2005) 3} [\href{https://arxiv.org/abs/hep-ph/0504242}{{\tt
  hep-ph/0504242}}].

\bibitem{Moch:2007gx}
S.~Moch and M.~Rogal, \emph{{Charged current deep-inelastic scattering at three
  loops}},
  \href{https://dx.doi.org/10.1016/j.nuclphysb.2007.05.008}{\emph{Nucl. Phys.
  B} {\bf 782} (2007) 51} [\href{https://arxiv.org/abs/0704.1740}{{\tt
  arXiv:0704.1740}}].

\bibitem{Moch:2007rq}
S.~Moch, M.~Rogal and A.~Vogt, \emph{{Differences between charged-current
  coefficient functions}},
  \href{https://dx.doi.org/10.1016/j.nuclphysb.2007.09.022}{\emph{Nucl. Phys.
  B} {\bf 790} (2008) 317} [\href{https://arxiv.org/abs/0708.3731}{{\tt
  arXiv:0708.3731}}].

\bibitem{Moch:2008fj}
S.~Moch, J.A.M.~Vermaseren and A.~Vogt, \emph{{Third-order QCD corrections to
  the charged-current structure function F(3)}},
  \href{https://dx.doi.org/10.1016/j.nuclphysb.2009.01.001}{\emph{Nucl. Phys.
  B} {\bf 813} (2009) 220} [\href{https://arxiv.org/abs/0812.4168}{{\tt
  arXiv:0812.4168}}].

\bibitem{Velizhanin:2011es}
V.N.~Velizhanin, \emph{{Four loop anomalous dimension of the second moment of
  the non-singlet twist-2 operator in QCD}},
  \href{https://dx.doi.org/10.1016/j.nuclphysb.2012.03.006}{\emph{Nucl. Phys.
  B} {\bf 860} (2012) 288} [\href{https://arxiv.org/abs/1112.3954}{{\tt
  arXiv:1112.3954}}].

\bibitem{Velizhanin:2014fua}
V.N.~Velizhanin, \emph{{Four-loop anomalous dimension of the third and fourth
  moments of the nonsinglet twist-2 operator in QCD}},
  \href{https://dx.doi.org/10.1142/S0217751X20501997}{\emph{Int. J. Mod. Phys.
  A} {\bf 35} (2020) 2050199} [\href{https://arxiv.org/abs/1411.1331}{{\tt
  arXiv:1411.1331}}].

\bibitem{Ruijl:2016pkm}
B.~Ruijl, T.~Ueda, J.A.M.~Vermaseren, J.~Davies and A.~Vogt, \emph{{First
  Forcer results on deep-inelastic scattering and related quantities}},
  \href{https://dx.doi.org/10.22323/1.260.0071}{\emph{PoS} {\bf LL2016} (2016)
  071} [\href{https://arxiv.org/abs/1605.08408}{{\tt arXiv:1605.08408}}].

\bibitem{Moch:2017uml}
S.~Moch, B.~Ruijl, T.~Ueda, J.A.M.~Vermaseren and A.~Vogt, \emph{{Four-Loop
  Non-Singlet Splitting Functions in the Planar Limit and Beyond}},
  \href{https://dx.doi.org/10.1007/JHEP10(2017)041}{\emph{JHEP} {\bf 10} (2017)
  041} [\href{https://arxiv.org/abs/1707.08315}{{\tt arXiv:1707.08315}}].

\bibitem{Moch:2018wjh}
S.~Moch, B.~Ruijl, T.~Ueda, J.A.M.~Vermaseren and A.~Vogt, \emph{{On quartic
  colour factors in splitting functions and the gluon cusp anomalous
  dimension}},
  \href{https://dx.doi.org/10.1016/j.physletb.2018.06.017}{\emph{Phys. Lett. B}
  {\bf 782} (2018) 627} [\href{https://arxiv.org/abs/1805.09638}{{\tt
  arXiv:1805.09638}}].

\bibitem{Moch:2021qrk}
S.~Moch, B.~Ruijl, T.~Ueda, J.A.M.~Vermaseren and A.~Vogt, \emph{{Low moments
  of the four-loop splitting functions in QCD}},
  \href{https://dx.doi.org/10.1016/j.physletb.2021.136853}{\emph{Phys. Lett. B}
  {\bf 825} (2022) 136853} [\href{https://arxiv.org/abs/2111.15561}{{\tt
  arXiv:2111.15561}}].

\bibitem{Gracey:1994nn}
J.A.~Gracey, \emph{{Anomalous dimension of nonsinglet Wilson operators at
  $O(1/n_f)$ in deep inelastic scattering}},
  \href{https://dx.doi.org/10.1016/0370-2693(94)90502-9}{\emph{Phys. Lett. B}
  {\bf 322} (1994) 141} [\href{https://arxiv.org/abs/hep-ph/9401214}{{\tt
  hep-ph/9401214}}].

\bibitem{Davies:2016jie}
J.~Davies, A.~Vogt, B.~Ruijl, T.~Ueda and J.A.M.~Vermaseren, \emph{{Large-$n_f$
  contributions to the four-loop splitting functions in QCD}},
  \href{https://dx.doi.org/10.1016/j.nuclphysb.2016.12.012}{\emph{Nucl. Phys.
  B} {\bf 915} (2017) 335} [\href{https://arxiv.org/abs/1610.07477}{{\tt
  arXiv:1610.07477}}].

\bibitem{Buras:1979yt}
A.J.~Buras, \emph{{Asymptotic Freedom in Deep Inelastic Processes in the
  Leading Order and Beyond}},
  \href{https://dx.doi.org/10.1103/RevModPhys.52.199}{\emph{Rev. Mod. Phys.}
  {\bf 52} (1980) 199}.

\bibitem{Moch:2004pa}
S.~Moch, J.A.M.~Vermaseren and A.~Vogt, \emph{{The Three loop splitting
  functions in QCD: The Nonsinglet case}},
  \href{https://dx.doi.org/10.1016/j.nuclphysb.2004.03.030}{\emph{Nucl. Phys.
  B} {\bf 688} (2004) 101} [\href{https://arxiv.org/abs/hep-ph/0403192}{{\tt
  hep-ph/0403192}}].

\bibitem{Vogt:2004mw}
A.~Vogt, S.~Moch and J.A.M.~Vermaseren, \emph{{The Three-loop splitting
  functions in QCD: The Singlet case}},
  \href{https://dx.doi.org/10.1016/j.nuclphysb.2004.04.024}{\emph{Nucl. Phys.
  B} {\bf 691} (2004) 129} [\href{https://arxiv.org/abs/hep-ph/0404111}{{\tt
  hep-ph/0404111}}].

\bibitem{Nogueira:1991ex}
P.~Nogueira, \emph{{Automatic Feynman graph generation}},
  \href{https://dx.doi.org/10.1006/jcph.1993.1074}{\emph{J. Comput. Phys.} {\bf
  105} (1993) 279}.

\bibitem{vanRitbergen:1998pn}
T.~van~Ritbergen, A.N.~Schellekens and J.A.M.~Vermaseren, \emph{{Group theory
  factors for Feynman diagrams}},
  \href{https://dx.doi.org/10.1142/S0217751X99000038}{\emph{Int. J. Mod. Phys.
  A} {\bf 14} (1999) 41} [\href{https://arxiv.org/abs/hep-ph/9802376}{{\tt
  hep-ph/9802376}}].

\bibitem{Larin:1996wd}
S.A.~Larin, P.~Nogueira, T.~van~Ritbergen and J.A.M.~Vermaseren, \emph{{The
  Three loop QCD calculation of the moments of deep inelastic structure
  functions}},
  \href{https://dx.doi.org/10.1016/S0550-3213(97)80038-7}{\emph{Nucl. Phys. B}
  {\bf 492} (1997) 338} [\href{https://arxiv.org/abs/hep-ph/9605317}{{\tt
  hep-ph/9605317}}].

\bibitem{Tkachov:1981wb}
F.V.~Tkachov, \emph{{A Theorem on Analytical Calculability of Four Loop
  Renormalization Group Functions}},
  \href{https://dx.doi.org/10.1016/0370-2693(81)90288-4}{\emph{Phys. Lett. B}
  {\bf 100} (1981) 65}.

\bibitem{Chetyrkin:1981qh}
K.G.~Chetyrkin and F.V.~Tkachov, \emph{{Integration by Parts: The Algorithm to
  Calculate beta Functions in 4 Loops}},
  \href{https://dx.doi.org/10.1016/0550-3213(81)90199-1}{\emph{Nucl. Phys. B}
  {\bf 192} (1981) 159}.

\bibitem{Ruijl:2017cxj}
B.~Ruijl, T.~Ueda and J.A.M.~Vermaseren, \emph{{Forcer, a FORM program for the
  parametric reduction of four-loop massless propagator diagrams}},
  \href{https://dx.doi.org/10.1016/j.cpc.2020.107198}{\emph{Comput. Phys.
  Commun.} {\bf 253} (2020) 107198}
  [\href{https://arxiv.org/abs/1704.06650}{{\tt arXiv:1704.06650}}].

\bibitem{Vermaseren:2000nd}
J.A.M.~Vermaseren, \emph{{New features of FORM}},
  \href{https://arxiv.org/abs/math-ph/0010025}{{\tt math-ph/0010025}}.

\bibitem{Kuipers:2012rf}
J.~Kuipers, T.~Ueda, J.A.M.~Vermaseren and J.~Vollinga, \emph{{FORM version
  4.0}}, \href{https://dx.doi.org/10.1016/j.cpc.2012.12.028}{\emph{Comput.
  Phys. Commun.} {\bf 184} (2013) 1453}
  [\href{https://arxiv.org/abs/1203.6543}{{\tt arXiv:1203.6543}}].

\bibitem{Ruijl:2017dtg}
B.~Ruijl, T.~Ueda and J.~Vermaseren, \emph{{FORM version 4.2}},
  \href{https://arxiv.org/abs/1707.06453}{{\tt arXiv:1707.06453}}.

\bibitem{Tentyukov:2007mu}
M.~Tentyukov and J.A.M.~Vermaseren, \emph{{The Multithreaded version of FORM}},
  \href{https://dx.doi.org/10.1016/j.cpc.2010.04.009}{\emph{Comput. Phys.
  Commun.} {\bf 181} (2010) 1419}
  [\href{https://arxiv.org/abs/hep-ph/0702279}{{\tt hep-ph/0702279}}].

\bibitem{tHooft:1973mfk}
G.~'t~Hooft, \emph{{Dimensional regularization and the renormalization group}},
  \href{https://dx.doi.org/10.1016/0550-3213(73)90376-3}{\emph{Nucl. Phys. B}
  {\bf 61} (1973) 455}.

\bibitem{Mankiewicz:1997gz}
L.~Mankiewicz, M.~Maul and E.~Stein, \emph{{Perturbative part of the nonsinglet
  structure function $F_2$ in the large $n_f$ limit}},
  \href{https://dx.doi.org/10.1016/S0370-2693(97)00568-6}{\emph{Phys. Lett. B}
  {\bf 404} (1997) 345} [\href{https://arxiv.org/abs/hep-ph/9703356}{{\tt
  hep-ph/9703356}}].

\bibitem{Gracey:1995aj}
J.A.~Gracey, \emph{{Large $n_f$ methods for computing the perturbative
  structure of deep inelastic scattering}},  in \emph{{4th International
  Workshop on Software Engineering and Artificial Intelligence for High-energy
  and Nuclear Physics}}, 9, 1995.
\newblock \href{https://arxiv.org/abs/hep-ph/9509276}{{\tt hep-ph/9509276}}.

\bibitem{Das:2019btv}
G.~Das, S.O.~Moch and A.~Vogt, \emph{{Soft corrections to inclusive
  deep-inelastic scattering at four loops and beyond}},
  \href{https://dx.doi.org/10.1007/JHEP03(2020)116}{\emph{JHEP} {\bf 03} (2020)
  116} [\href{https://arxiv.org/abs/1912.12920}{{\tt arXiv:1912.12920}}].

\bibitem{Moch:2009hr}
S.~Moch and A.~Vogt, \emph{{On non-singlet physical evolution kernels and
  large-x coefficient functions in perturbative QCD}},
  \href{https://dx.doi.org/10.1088/1126-6708/2009/11/099}{\emph{JHEP} {\bf 11}
  (2009) 099} [\href{https://arxiv.org/abs/0909.2124}{{\tt arXiv:0909.2124}}].

\bibitem{Ajjath:2020sjk}
A.H.~Ajjath, P.~Mukherjee, V.~Ravindran, A.~Sankar and S.~Tiwari, \emph{{On
  next to soft threshold corrections to DIS and SIA processes}},
  \href{https://dx.doi.org/10.1007/JHEP04(2021)131}{\emph{JHEP} {\bf 04} (2021)
  131} [\href{https://arxiv.org/abs/2007.12214}{{\tt arXiv:2007.12214}}].

\bibitem{Vogt:2012gb}
A.~Vogt, C.H.~Kom, N.A.~Lo~Presti, G.~Soar, A.A.~Almasy, S.~Moch et~al.,
  \emph{{Progress on double-logarithmic large-x and small-x resummations for
  (semi-)inclusive hard processes}},
  \href{https://dx.doi.org/10.22323/1.151.0004}{\emph{PoS} {\bf LL2012} (2012)
  004} [\href{https://arxiv.org/abs/1212.2932}{{\tt arXiv:1212.2932}}].

\bibitem{Davies:2022ofz}
J.~Davies, C.H.~Kom, S.~Moch and A.~Vogt, \emph{{Resummation of small-x double
  logarithms in QCD: inclusive deep-inelastic scattering}},
  \href{https://arxiv.org/abs/2202.10362}{{\tt arXiv:2202.10362}}.

\bibitem{Baikov:2010je}
P.A.~Baikov, K.G.~Chetyrkin and J.H.~K{\"u}hn, \emph{{Adler Function, Bjorken
  Sum Rule, and the Crewther Relation to Order $\alpha^4_s$ in a General Gauge
  Theory}},
  \href{https://dx.doi.org/10.1103/PhysRevLett.104.132004}{\emph{Phys. Rev.
  Lett.} {\bf 104} (2010) 132004} [\href{https://arxiv.org/abs/1001.3606}{{\tt
  arXiv:1001.3606}}].

\bibitem{Baikov:2012zn}
P.A.~Baikov, K.G.~Chetyrkin, J.H.~K{\"u}hn and J.~Rittinger, \emph{{Adler
  Function, Sum Rules and Crewther Relation of Order $\mathcal{O}(\alpha^4_s)$:
  the Singlet Case}},
  \href{https://dx.doi.org/10.1016/j.physletb.2012.06.052}{\emph{Phys. Lett. B}
  {\bf 714} (2012) 62} [\href{https://arxiv.org/abs/1206.1288}{{\tt
  arXiv:1206.1288}}].

\bibitem{Herzog:2018kwj}
F.~Herzog, S.~Moch, B.~Ruijl, T.~Ueda, J.A.M.~Vermaseren and A.~Vogt,
  \emph{{Five-loop contributions to low-N non-singlet anomalous dimensions in
  QCD}}, \href{https://dx.doi.org/10.1016/j.physletb.2019.01.060}{\emph{Phys.
  Lett. B} {\bf 790} (2019) 436} [\href{https://arxiv.org/abs/1812.11818}{{\tt
  arXiv:1812.11818}}].

\bibitem{Retey:2000nq}
A.~Retey and J.A.M.~Vermaseren, \emph{{Some higher moments of deep inelastic
  structure functions at next-to-next-to-leading order of perturbative QCD}},
  \href{https://dx.doi.org/10.1016/S0550-3213(01)00149-3}{\emph{Nucl. Phys. B}
  {\bf 604} (2001) 281} [\href{https://arxiv.org/abs/hep-ph/0007294}{{\tt
  hep-ph/0007294}}].

\bibitem{vanNeerven:2000wp}
W.L.~van~Neerven and A.~Vogt, \emph{{Improved approximations for the three loop
  splitting functions in QCD}},
  \href{https://dx.doi.org/10.1016/S0370-2693(00)00953-9}{\emph{Phys. Lett. B}
  {\bf 490} (2000) 111} [\href{https://arxiv.org/abs/hep-ph/0007362}{{\tt
  hep-ph/0007362}}].

\bibitem{vanNeerven:2001pe}
W.L.~van~Neerven and A.~Vogt, \emph{{Nonsinglet structure functions beyond the
  next-to-next-to-leading order}},
  \href{https://dx.doi.org/10.1016/S0550-3213(01)00158-4}{\emph{Nucl. Phys. B}
  {\bf 603} (2001) 42} [\href{https://arxiv.org/abs/hep-ph/0103123}{{\tt
  hep-ph/0103123}}].

\bibitem{MRUVV-tba}
S.~Moch, B.~Ruijl, T.~Ueda, J.A.M.~Vermaseren and A.~Vogt, \emph{{to appear}}.

\end{thebibliography}

\providecommand{\href}[2]{#2}\begingroup\raggedright\endgroup


\end{document}